# ALGORITHM FOR SPECTRAL RESPONSE ANALYSIS OF SUPERCONDUCTING MICROWAVE TRANSMISSION-LINE RESONATOR

**Muhammad Hanif** *
*DISAT, Politecnico di Torino, Torino 10129, Italy.
muhammad.hanif(at)polito(dot)it

***ABSTRACT:****It has always been a challenge for researchers to efficiently and accurately post process experimental data which is distorted by the noise. Superconducting microwave devices e.g. resonators, directional filters, beam-splitters etc. operate at frequency of several GHz to THz and temperatures well below critical temperature ($T_c$) with few exceptions like transition edge sensors where devices are operated at temperatures close to $T_c$. These devices are measured usually with vector network analyzer in terms of scattering parameters.Vector network analyzer has a base on mature technology which has seen many developments after its first use, more than half a century ago. With all the break troughs in electronics and cryogenics, the measurement tools still have three kinds of errors i.e. systematic, drift and random errors causing headache to the researchers and scientists at times. Two of these errors, systematic and drift, can easily be removed from the measurements taken with VNA. However, random errors are not easy to address and remove due to their unpredictability and randomness. In this manuscript we will present an algorithm to post process experimental data to cope with measurements that have been corrupted or useful spectral response is buried in spurious signal. We have developed a robust and efficient algorithm, implemented in MATLAB, to detect peaks in spectral response, remove baseline andfinally estimate parameters of two-port superconductor resonator using an Improved Nelder-Mead Method for unconstrained multidimensional least square minimization.The algorithm has been successfully tested and verified by processing spectral response of half wavelength microwave transmission-line resonator successfully isolating resonator response from noisy background. We were able to compute loaded quality factor, resonance frequency from response data with high reproducibility even from those experimental data sets where resonance spikes were hardly visible.*

## I. INTRODUCTION

We have introduced a Robust Peak Analysis Algorithm (RPAA) which provide a mean for quantitative baseline estimation numerically from spectra which is a mixture of sharp features (high frequency peaks) and continuous, slowly varying (low frequency) baseline. Though, we have used RPAA to post process spectral response of superconducting microwave resonators but it can be employed to any spectroscopic studies owing to its robustness and general applicability.

There are, certainly, various methods to remove the baseline from spectra, for the reason that the problem of noisy spectral data is omnipresent within spectroscopy. It is usual practice, when high accuracy is not desirable, to remove to the baseline modestly 'by eye'. Inherently, this method doesn't have anything wrong with it; however, for a mundane task of baseline estimation, numerical tools are valuable which not only minimize the user interference but also ensure reproducibility. The knack to deter the peak intensity quantitatively and peak shape information is, generally, restricted by the underlying baseline deterrence and under these circumstances numerical techniques are clearly unavoidable to estimate and remove baseline.

There is sparse work addressing post processing of resonance data of microwavetransmission-line resonators (MTR) recorded by VNA. The available literature on peak data analysis is dispersed across diverse research fields including analytical chemistry [1], nuclear physics [2], X-ray and mass spectroscopy [3], nuclear magnetic resonance (NMR) spectroscopy [4] and medical research [5]. In reference [1], some general approaches for baselineestimation are comprehensively described. A widely used approach for baseline estimation is to establish a function by taking into account only those regions of spectrum comprising baseline solely, and then, interpolating the estimated baseline function in the regions of spectrum where peaks exist. This approach is called 'automated peak rejection algorithm' [4]. Other well-known methods include digital filtering [3], principle component analysis [6] and maximum entropy methods [1].

The RPAA employs robust local regression for baseline estimation, extract peak shape parameters efficiently fitting a priori known peak model using Improved Nelder Mead Simplex (INMS) [7] method in least square sense. Briefly, the attributes of the RPAA are;

1. It relies on local regression method which takes into account measurement errors effectively.
2. It has flexibility to address variety of peak analysis problems.
3. The human interventions are reduced in contrast to other known methods.

The manuscript is organized as; in section II, details on baseline estimation using robust local regression method are given. Section III contains INMS method details used to estimate the peak shape features. Finally, results and discussion are given in section IV followed by conclusion.

## II. BASELINE ESTIMATION

We start by differentiating local and robust estimation from well-known least square fitting. In particular, our focus will be on locally weighted scatter smoother (LOWESS) introduces by Cleveland [8] which is employed for baseline estimation in RPAA.

Let a predictor variable *p* corresponding to a response *R*. We can write it as a data set $(p_i, R_i)$, such that i=1, 2,…n. Relationship between pi and $R_i$ may be established by regressive fitting of data. The difference of two, data and regressive curve, is treated as noise.

$$R_i = g(p_i) + \epsilon_i \ (i = 1, 2, \ldots, n), \qquad (1)$$

Second term on right side is known as random error and often considered independent and symmetrically distributed i.e. they have zero mean and variance σ2. Least squares approach remains most successful estimation technique in case regression curve has known form up to unknown parameters $\theta = (\theta_1, \ldots, \theta_q)T$.

$$\hat{\theta} = \arg\min_\theta \sum_{i=1}^{n} [R_i - g(p_i:\theta)]^2, \qquad (2)$$

The functional form of regression curve, g in Eq. (1), is not always predictable. However, if g can be considered smooth function, locally linear regression model can also be applied for estimation. In this method, regression curve *g(p)* is considered to be linear in sufficiently small vicinity of a given point $p_0$ and baseline is then approximated by applying least-squares technique. In addition to local least-squares problem, a weight scheme can be incorporated through which influence of data points can be altered in proportion to their distance from $p_0$.

$$\hat{\theta} = \arg\min_\theta \sum_{i=1}^{n} K\left(\frac{p_i - p_0}{h}\right)[y_i - \{\theta_0 + \theta_1(x_i - x_0)\}]^2, \qquad (3)$$

Here, $K\left(\frac{p_i - p_0}{h}\right)$ is kernel weight of $p_i$ which is nearly zero far from$p_0$. The precise form of *K* marginally affects the estimator's performance, both theoretically and empirically [9]. Thus, weight kernel is chosen on the basis of computational ease. Robust local regression uses tricube weight function which smoothly descends to zero away from the neighborhood of $p_0$ defined by $p_0 \pm h$, contrary to Gaussian that normally has nonzero tails beyond$p_0 \pm h$.

$$K(u) = 0.9.*(1 - |u|^3)^3, \qquad (4)$$

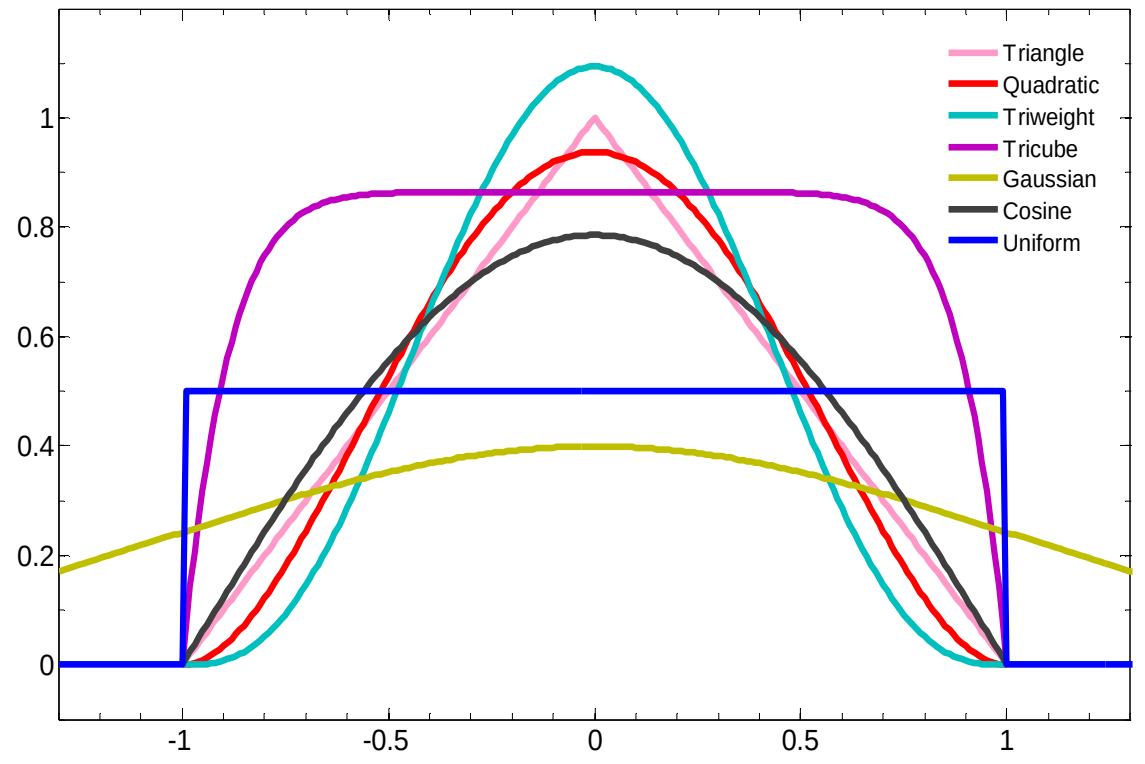

Figure 1.Different weight kernels in arbitrary coordinate system.

Figure 1 shows a comparison of several weight kernels which can be used in accordance with the needs of the problem. Usually, RF response of superconducting resonators measured with VNA contains high frequency fluctuations (need to be filtered out) superimposed on to low frequency smooth background, which we would refer to as baseline from now on. The baseline is estimated aftersmoothening of high frequency kinks in the measured signal. Tricube kernel is a natural selection in current situation as it effectively incorporate neighborhood points and give less weightage to outliers ensuring that the peak of interest is not affected by the process of baseline estimation.

The value of *h* in equation 3is very crucial, as if its value is enormously small, the resultant estimation is effectively interpolation of data points, while, very large value of h results in, essentially, a globally linear regression solution.

Local regression method given by equation 3 is sensitive to outliers and make it difficult to distinguish between intrinsic features of data i.e. peaks or jumps and systematic outliers. Onepossible approach, however, is to evaluate local least-squares (equation 3)to estimate *g(p)* initially and assign residuals of this fit to robustness weights, $w_r(p_i)$, determined by Tukey's equation (equation 6), to every point $(p_i, R_i)$ such that points with small residual get larger weights and vice versa [10]. The regression curve can then be calculated by weighted least square fit (equation 5) and repeated iteratively to convergence with $w_r(p_i)$ always determined from previous iteration.

$$\hat{\theta} = \arg\min_\theta \sum_{i=1}^{n} w_r(p_i) K\left(\frac{p_i - p_0}{h}\right)[y_i - \{\theta_0 + \theta_1(x_i - x_0)\}]^2 \qquad (5)$$

$$w_r(p_i) = \left\{1 - \left(\frac{r_i}{b}\right)^2\right\}^2 \qquad (6)$$

Where$r_i = \frac{[y_i - g(p_i)]}{\sigma}$. The parameter *b*in equation 6 defines degree of process robustness i.e. how strongly the fit is influenced by outliers. A careful analysis is required to choose *b*such that useful data (Peak) is not smoothened out. We have made our coding flexible in selecting valueof *b* (Cleveland chose *b*=4.05) which we call *R*-Parameter which is inversely proportional to *b*. Implication of this change is; it makes LOESS method more robust, efficient and allowing very noisy data to be processed. This is the only parameter where user can intervene to change the degree of robustness to meet the desired accuracy in baseline construction i.e. estimation of $g(p_i)$. Further details on R-Parameter and effect on $g(p_i)$estimation will be discussed in section-IV.

In order to implement the LOESS method, we need to specify scale parameter *(σ)* which conveniently can be taken as experimental noise. In some experiments, we do know *σ* a priori. However, if it is not known, we can estimate it using the median of absolute values (MAV) of the residuals [10].

$$\sigma_{MAV} = \frac{Mdn[|y_i - g(p_i)|]}{0.6745} \qquad (7)$$

We can summarize the baseline estimation process as under;

- If the $g(p_i)$,calculated using equation 3,4, of a point $p_i$ is smaller than the weight kernel, then the intensity of corresponding point on $g(p_i)$ equals the intensity of $p_i$.
- If the intensity of a point is larger than or equal to the weight kernel, then the intensity of correspondingpoint on $g(p_i)$ equals $g(p_i)$.

- Baseline is obtained by applying local regressionfitting to the predictor(equation 5, 6).
- Process can be iterated to achieve convergence.

## III. INMS METHOD

Once we have clean, baseline rectified, RF response data of planar resonators we are required to estimate resonator characteristics i.e. quality factor *(Q)*, resonance frequency *($f_0$)* etc. Resonator response is usually characterized by a Lorentziancurve especially when resonance responses are very weak.

In this section, we shall briefly describe INMS method for unconstrained parametric optimization of modified Lorentzian function given by [11];

$$S_{21} = \frac{a_1}{\sqrt{1+a_2^2\left(\frac{f}{a_3}-\frac{a_3}{f}\right)}} + a_4 f + a_5 \qquad (8)$$

Here $a_1$is amplitude at resonance, $a_2$ corresponds to $Q$ of resonator, $a_3$represents $f_0$, $a_4$ is slope and $a_5$ offset.$a_i$; i=1-5, are parameters we are aiming to optimize using INMS. First, we will summarize Nelder Mead following original paper [12] followed by modifications by[7];

1. Get $\alpha$ (contraction), $\beta$ (reflection) and $\gamma$ (expansion) coefficients, chose initial simplex with random vertices $x_0, x_1, \ldots, x_n$ alsocompute corresponding function values $f_0, f_1, \ldots, f_n$.
2. Sort the vertices so that $f_0, f_1, \ldots, f_n$ are in ascending order.
3. Compute the reflection point$x_r$, $f_r$.
4. if $f_0 > f_r$:
   - 4.1. calculate the extended point$x_e$, $f_e$;
   - 4.2. if $f_e < f_0$, substitute the worst point with extendedpoint i.e. $x_n = x_e$, $f_n = f_e$;
   - *4.3.* if$f_e > f_0$, substitute the worst point with reflectedpoint $x_n = x_r$, $f_n = f_r$.
5. if $f_0 < f_r$
   - *5.1.* if$f_r < f_i$,substitute the worst point with reflected point $x_n = x_r$, $f_n = f_r$.
   - 5.2. if $f_r > f_i$:
     - *5.2.1.* if $f_r > f_n$: calculate the contracted point $x_c$, $f_c$;
     - 5.2.2. if $f_c > f_n$ then contract the simplex;
     - 5.2.3. if $f_c < f_n$, substitute the worst point with contracted point $x_n = x_c$, $f_n = f_c$;
   - *5.3.* if$f_r < f_n$: substitute the worst point withreflected point $x_n = x_r$, $f_n = f_r$.
6. if the stopping criteria not met, the algorithm will continue at 2.

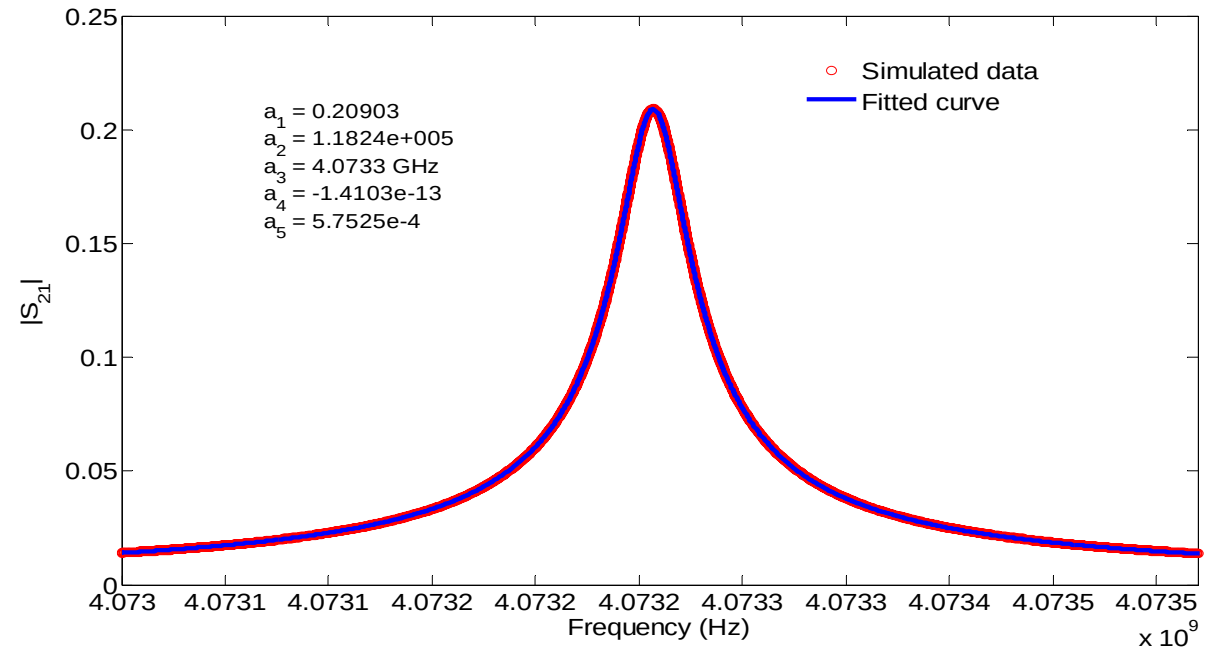

Figure 2. Simulated forward scattering of a $\lambda/2$ microwave transmission line resonators (MTR) and Lorentz fit.

Nelder-Mead simplex method, as can be seen, is very robust and computation efficient method to solve unconstrained problems without derivative calculation. The major drawback, however, of NMS method is that it may not define its moving directions well enough just by simple geometrical movements in high dimensional cases.

To maintain this simplicity, one quasi gradient method presented by Nam Pham et al. for gradient approximation of a function. This method involves an extra point created from a simplex to approximate gradients. It approximates gradients of a (n+1) dimensional plane created from a geometrical simplex. By approximating gradients of the plane, gradients in the vicinity of a simplex can be approximated. A comprehensive detail of INMS modification steps can be found in reference [7].

INMS is essentially a direct search iterative method whose efficiency of convergence largely depends upon initial guess of parameters needed to be optimized. Now we will make a note on initial guess parameters involved in equation 8.Let *y(x)* be the real part of complex RF response of resonator where x is frequency and baseline is estimated and subtracted according to method described in Section I. Initial guess for *offset*is evaluated as;

$$a_5 = \frac{y_{start}x_{end} - x_{start}y_{end}}{x_{end} - x_{start}} \qquad (9)$$

In equation 9, $y_{start}$ and $y_{end}$ are averages of first and last ten data points of $y$, respectively. Similarly,$x_{start}$ and $x_{end}$are averages of first and last ten points of $x$,respectively. The slope can be approximated as;

$$a_4 = \frac{y_{end} - y_{start}}{x_{end} - x_{start}} \qquad (10)$$

Peak height $a_1$ and position of peak,$f_0 = a_3$, can be easily found using MATLAB built-in functions.

*[$a_1$, index]= max(y);* (11)

*$a_3$=x(index);* (12)

The only parameter left so far is $a_2$ which corresponds to $Q$ of the resonator which is equal to $f_0/fwhm$. The fwhm is full width at half maximum of the peak which can be calculated using MATLAB function *fwhm.m* and we get $a_2$ as follows,

*$a_2$=$a_3$/fwhm* (13)

We can summarize the algorithm as follows;

1. Use equations1-7 for baseline estimation and subtract the baseline from original measured data.
2. Use equation 9-13 to get initial values for model (equation 8) parameters.
3. Fit the model (equation 8) to refined experimental data from first step using INMS method in least square approach.

## 4. IV. IMPLENTATION AND RESULTS

In this section we will see how the robust technique, described in previous section, work with some experimentally recorded data. We have implemented the RPAA in MATLAB which is very useful owing to its built-in capabilities of direct data acquisition from network analyser, analyse data immediately without hassle of saving and importing into MATLAB at a later time, making spectral analysis straight forward. Further, programming capabilities of MATLAB allows automated data analysis of huge amount of data very quickly.

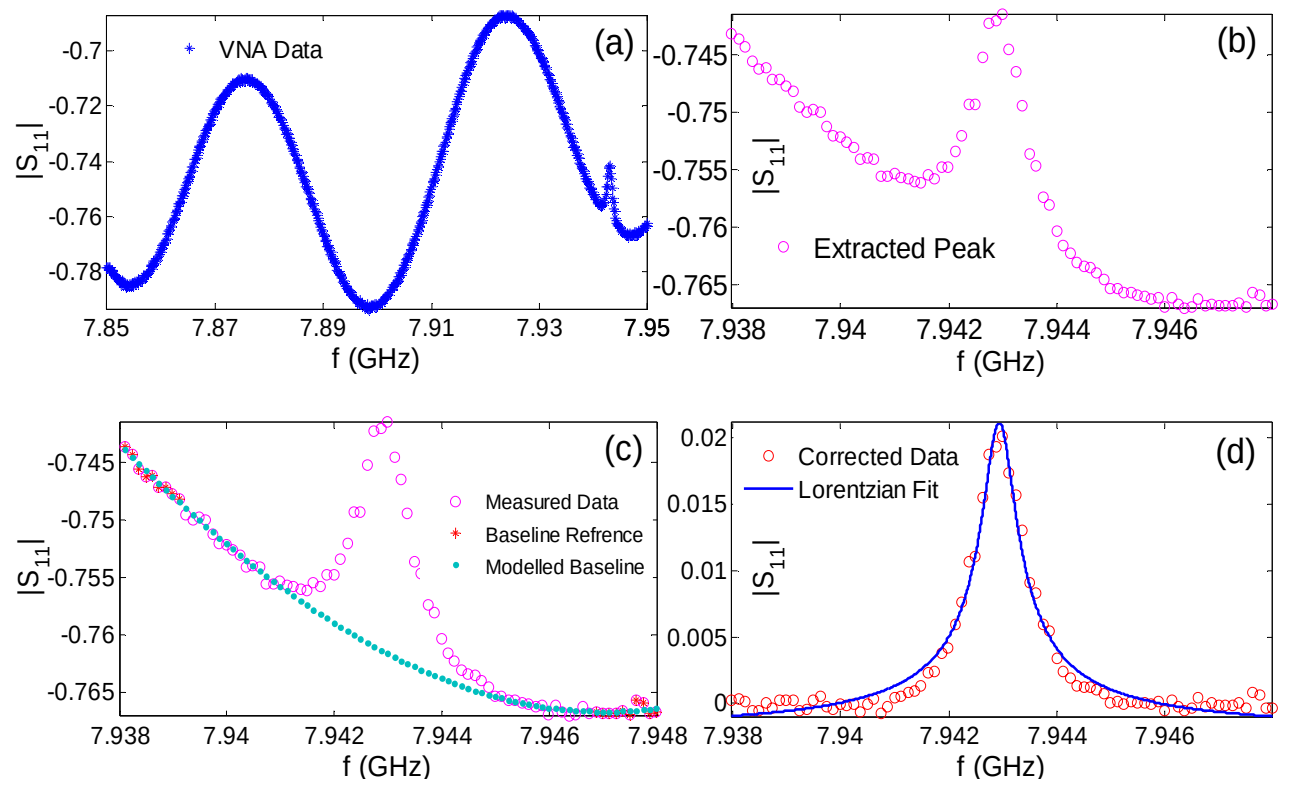

Figure 3. Measured scattering parameter: a) shows the original data b) Recovered peak using Peak finding algorithm (PFA) described in section-2 c) Modelled baseline fitting to extracted peak and d) shows Lorentzian fit to corrected peak.

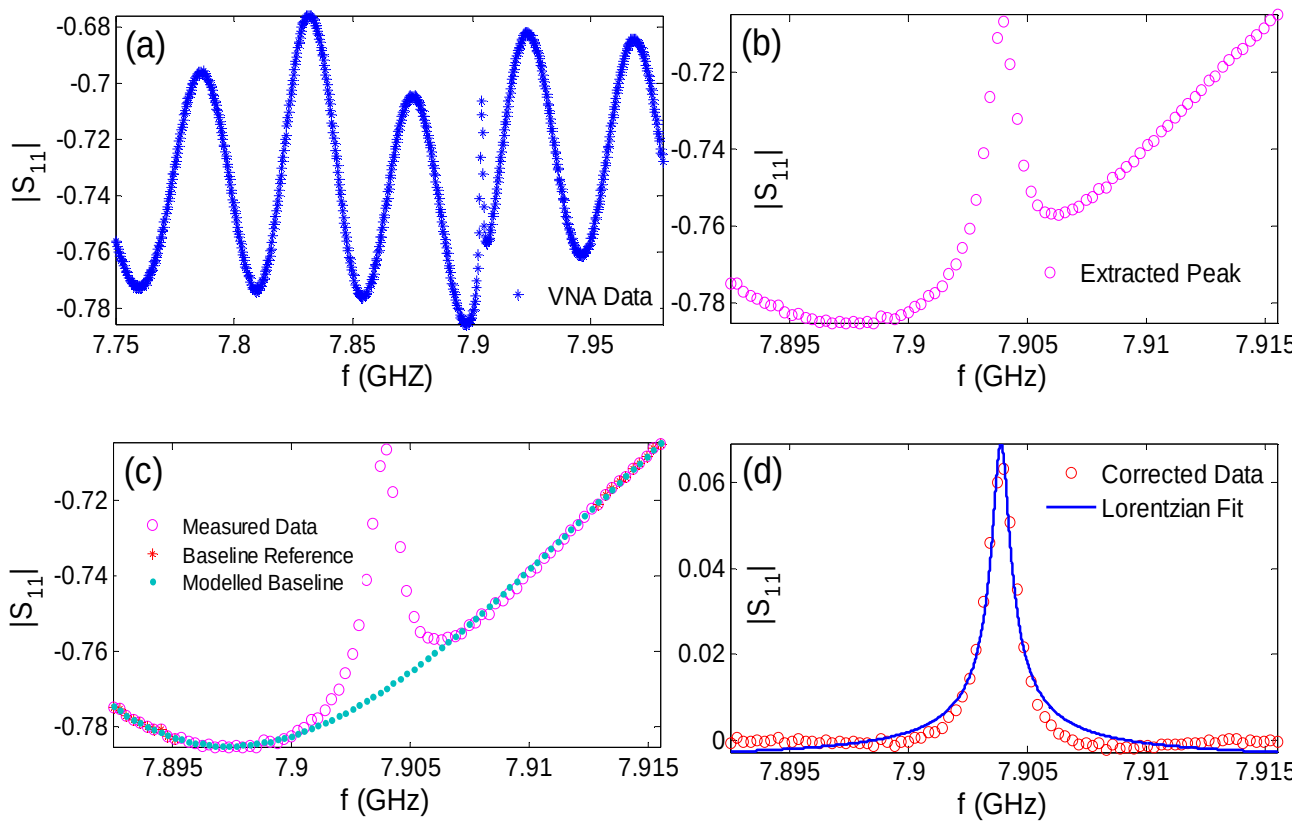

Figure 4.Measured scattering parameter of same device as in figure 3 but different experimental conditions resulting in different background noise.

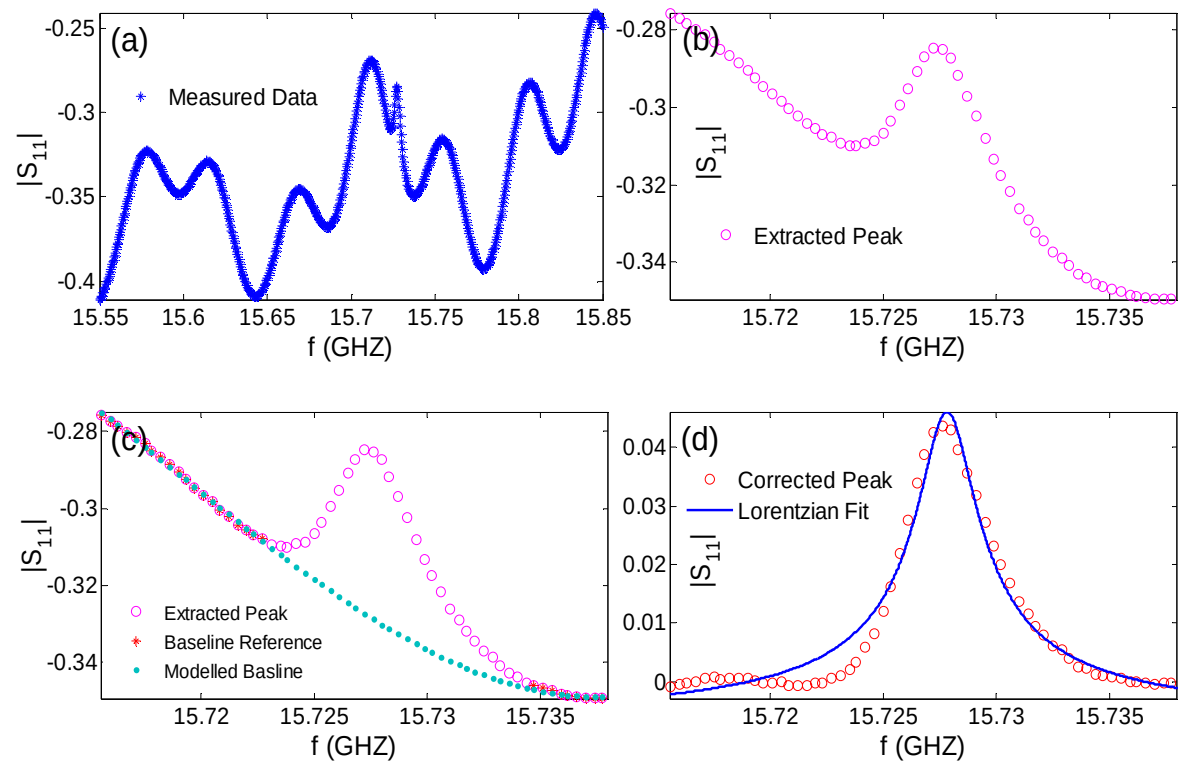

Figure 5.Case of even more worse background random noise.

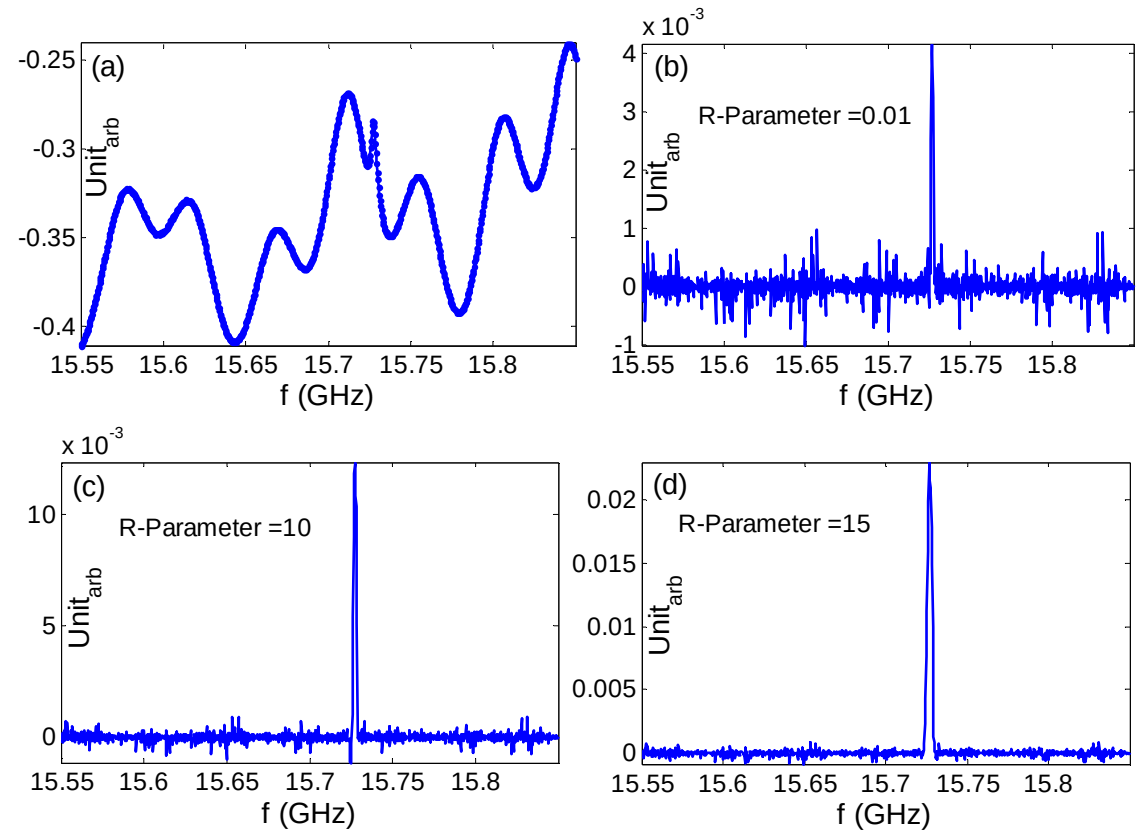

Figure 6: R-parameter value effect on extracted peak. a) Shows measured data, (b), (c), (d) showing recovered peak with R-Parameter values of 0.01, 10 and 15, respectively.

Algorithm has tendency to find peak even when global minimum/maximum is not evident in the displayed data. With the default parameters of robustness of RPAA;peak,which is 10% of any noisy signal can be retrieved. This can be tuned in accordance with data to be analysed. A higher value of robustness parameter will result in losing small peaks in data but making peak extraction quicker. However, a very small (<1%) value is not always beneficial as it may result in false peak detection (see figure 6-b) along with excessive computation cost and very large value (<10%) can mess up peak propertiesas can be seen in figure 6-c and 6-d where peak height and width has been altered effectively to give wrong information.

## V. CONCLUSION

In this manuscript, we have presented a method for post processing of a noisy peak data. We have successfully implemented the algorithm in MATLAB and tested on real time MTR device's response recorded by VNA. The success rate of peak retrieval we have attained is quite high (a spike of the order of 1 mV in slowly varying noisy background can be extracted. See figure 3) and peak parameter estimation accuracy is up to six decimal places which is default value to fit equation-8 to extracted peak using modified INMS described in section 3.

In the future work we will investigate the applicability of method in other areas of research where peak analysis is extensively required e.g. XRD, mass spectrometry, bio-informatics etc. Further, a comparative study of the method developed with other commercial tools of data analysis will also remain an area of interest.

**AKNOWLEDGEMENT:** We are highly grateful to Higher Education Commission (HEC) of Pakistan for providing funding for this research work to be conducted at Politecnico di Torino, Italy under UESTP-Italy project.

## REFERENCE

1. Phillips A. J. and Hamilton P. A.,*Anal. Chem.*, **68**(22), 4020-4025(1996).
2. Burgess DD. Nucl.*Instr. & Meth.inPhys. Res.,***221**(3), 593-599(1984).

3. Kneen MA, Annegard H. J. *Nucl.Instr. & Meth. in Phys. Res. B.,* **109/110**, 209-213(1996).
4. Dietrich W, RuKdel CH, Neumann M. *J.Magn.Reson.,* **91**(1), 1-11(1991).
5. Froning JN, Olson MD, Froelicher VF. *J. Electro-cardiol.,* **21**(sup.), 149-57(1988).
6. Balcerowska G, SiudaR.,*Vacuum,* **54**(1-4), 195-199(1999).
7. Nam Pham, Bogdan M. Wilamowski, *J.of Comp.,***3**(3), 55-63(2011).
8. Cleveland W. S.,*J. A. Stat. Assoc.,***74**(368), 829-836(1979).
9. Simonoff J. S., *Smoothing methods in statistics,* Chapter-4, (1996).
10. Andreas F. Ruckstuh, Matthew P. Jacobson, Robert W. Field, James A. Dodd, *J. of Quan. Spec.&Radi.Tran.*,**68**, 179-193(2001).
11. Paul J. Petersan and Steven M. Anlage.,*J. of App. Phys.*, **84**(6), 3392-3402(1998).
12. J. A. Nelder, R. Mead, *J. Comp.*, **7**(4), 308-313(1965).